\documentclass[twocolumns]{aa}  

\usepackage{graphicx}
\usepackage{txfonts}
\usepackage{natbib}
\usepackage{amstext}
\usepackage{color}

\begin{document}

   \title{Using the chromatic Rossiter-McLaughlin effect to probe the broadband signature in the optical transmission spectrum of HD 189733b}

   %\subtitle{I. Overviewing the $\kappa$-mechanism}

   \author{E. Di Gloria
          \inst{1}
          \and
          I. A. G. Snellen\inst{1}
        \and
        S. Albrecht\inst{2}
          }

   \institute{Leiden Observatory, Leiden University, Postbus 9513, 2300 RA, Leiden, The Netherlands\\
              \email{digloria@strw.leidenuniv.nl}
         \and
                Stellar Astrophysics Centre, Department of Physics and Astronomy, Aarhus University, Ny Munkegade 120, DK-8000 Aarhus C, Denmark
             }

   \date{Received 30 March 2015; accepted 23 June 2015}

% \abstract{}{}{}{}{} 
% 5 {} token are mandatory
 
  \abstract
  % context heading (optional)    
   {Transmission spectroscopy is a powerful technique for probing exoplanetary atmospheres. A successful ground-based observational method uses a differential technique that uses high-dispersion spectroscopy, but it only preserves narrow features in transmission spectra. Broadband features, such as the remarkable Rayleigh-scattering slope from possible hazes in the atmosphere of HD 189733b as observed by the Hubble Space Telescope, cannot be probed in this way.}
  % aims heading (mandatory)
   {Here we use the chromatic Rossiter-McLaughlin (RM) effect  to measure the Rayleigh-scattering slope in the transmission spectrum of HD 189733b with the aim to show that it can be effectively used to measure broadband transmission features. The amplitude of the RM effects depends on the effective size of the planet, and in the case of an atmospheric contribution therefore depends on the observed wavelength.}
  % methods heading (mandatory)
   {We analysed archival HARPS data of three transits of HD 189733b, 
covering a wavelength range of 400 to 700 nm. The radial velocity
(RV) time-series were determined for white light and for six wavelength bins 
each 50 nm wide, using the cross-correlation profiles as provided by the 
HARPS data reduction pipeline. The RM effect was first fitted to the white-light RV 
time series using the publicly available code AROME. 
The residuals to this best fit were subsequently subtracted from the RV time 
series of each wavelength bin, after which they were also fitted using the same code, leaving only the effective planet radius to vary.
}
  % results heading (mandatory)
   {We measured the slope in the transmission spectrum of HD 189733b at a $2.5\sigma$ significance. Assuming it is due to Rayleigh scattering and not caused by stellar activity, it would correspond to an atmospheric temperature, as set by the scale height, of $T = 2300 \pm 900 \mathrm{K}$, well in line with previously
obtained results.}
  % conclusions heading (optional), leave it empty if necessary 
   {Ground-based high-dispersion spectral observations can be 
used to probe broad-band features in the transmission spectra of 
extrasolar planets, such as the optical Rayleigh-scattering slope of 
HD 189733b, by using the chromatic Rossiter-McLaughlin effect. The precision achieved with HARPS per transit is about an order of magnitude lower than that with STIS on the Hubble Space Telescope. This method will be 
particularly interesting in conjunction with the new echelle 
spectrograph ESPRESSO, which currently is under construction for ESO’s Very Large Telescope, which will provide a gain in signal-to-noise ratio of about a factor $\text{}$4 compared to HARPS. This will be of great value because of the limited and uncertain future of the Hubble Space Telescope and because the future James Webb Space Telescope will not cover this wavelength regime.}

\titlerunning{Chromatic Rossiter-McLaughlin effect to probe Rayleigh scattering on HD 189733b}
   \maketitle

%
%________________________________________________________________

\section{Introduction}
Transmission spectroscopy has proven to be a powerful technique for probing the atmospheres of extrasolar planets \citep{Charbonneau02,Deming13}. Ground-based transmission spectroscopy is very challenging, even
though telescopes with much larger collecting areas can be used than from space. Observations from the ground are hampered by the fact that a target is always seen through a varying amount of atmosphere. Moreover, atmospheric circumstances are never perfectly stable over a timescale of a few hours. In addition, instrument stability is crucial, since ground-based observations are subject to changes in the gravity vector and/or field rotation. One way to mitigate these issues is by using ground-based high-dispersion spectroscopy. For example, detections of sodium and potassium have been presented in the optical\citep{Redfield08,Snellen08,Sing11-pot,Wood11,Jensen11,Zhoubayliss12,Wyttenbach15}, and detections of carbon monoxide and water in
the infrared \citep{Brogi12,Rodler12,Kok13,Birkby13,Lockwood14,Brogi14}.

Unfortunately, the high-dispersion spectroscopic technique is insensitive to broadband absorption features since only relatively narrow features can be probed because all large-scale structures in the spectra are removed during the data analysis to account for the telluric variability. Therefore, broadband features can only be probed from the ground using multi-object spectroscopy, which simultaneously observes the target and a number of reference stars to perform differential spectrophotometry to correct for atmospheric effects \citep{Bean10,Bean11}.

However, the most precise observations have been acquired from space. A particularly interesting broadband feature is the remarkable slope in the transmission spectrum of HD 189733b, as measured by \cite{Pont08} and \cite{Sing11} using the Hubble Space Telescope (HST). \cite{Pont13}, who discussed the UV and optical spectrum based on a total of eight transits observed with STIS and ACS on the HST, found the radius of the planet to decrease by $1.5 \%$ between $345$ nm and $775$ nm. \cite{Cavelier} and \cite{Pont13} interpreted this slope as caused by Rayleigh scattering by hazes at a temperature of $ \sim 1300 \ \mathrm{K}$, possibly increasing to $\sim 2000 \ \mathrm{K}$ at the shortest wavelengths probing the lowest pressure region in the atmosphere. It would be extremely valuable if there were another way to probe this feature from the ground.

\cite{Snellen04} devised a method for probing the varying planet size as function of wavelength of a transiting exoplanet in a different way. The technique makes use of the Rossiter-McLaughlin (RM) effect. If the orbital plane of the planet is aligned with the spin of the star, the transiting exoplanet will first block light from the approaching part and then from the receding part of the stellar surface because the
host star rotates. This effect results in a wobble in the radial velocity  (RV)of the star, first seen by \cite{Rossiter24} for eclipsing binaries, and recently observed for transiting planets in many systems \citep[e.g.][]{Queloz00}. To first order, the overall amplitude of this effect, $A_{\mathrm{RM}}$, is proportional to

\begin{equation}
A_{\mathrm{RM}} \propto V\sin i \left(\frac{R_{\rm{p}}^2}{R_{\rm{star}}^2 - R_{\rm{p}}^2} \right)
\label{rm_amplitude}
,\end{equation}

where $V\sin i$ is the projected rotation velocity of the star, $R_{\rm{p}}$ is the planet radius, and $R_{\rm{star}}$ is the stellar radius \citep{Haswell-book}. As can be seen, this amplitude depends on the effective radius of the planet, and since this (in case of observable transmission features) is wavelength dependent, subsequently the amplitude of the RM effect is also wavelength dependent and can be measured \citep{Snellen04}. The advantage of this method over conventional transmission spectroscopy measurements is that instead of using the intensity of off-transit spectra as a reference for in-transit spectra, it depends on the changes in the profiles of the stellar spectral lines in the same on-transit spectra. For ground-based observation this makes the technique less likely to be influenced by Earth atmospheric effects, which turned out to be challenging for the conventional approach, because line shapes are less influenced by telluric absorption. Numerical simulations have been performed by \cite{Dreizler09} to show the feasibility of this technique. Recently, \cite{Czesla12} used a technique based on very similar principles to study the chromosphere of active planet host-stars.

While \cite{Snellen04} advocated this technique to probe narrow absorption features, such as those from sodium, we now realize that this method can be very powerful in probing wide broadband features, which are very challenging to probe from the ground. In this paper we test this technique on the exoplanet HD 189733b using archival HARPS data. In Sect. 2 the archival data and initial data analysis are described, and in Sect. 3 the models used to fit the RM effect are presented. In Sects. 4 and 5 we present the fit to the data and discuss the final results. 

\section{HARPS data and radial velocity extraction}

HD 189733b is a hot Jupiter with a mass of $1.15 \ M_{\mathrm{Jupiter}}$ that orbits a K dwarf in $2.2$ days. The system parameters used for our analysis are presented in Table \ref{aromecoeff}.  It is one of the two most frequently studied and observed exoplanets, together with HD 209458b. Its atmosphere has been observed both in trasmission and during secondary eclipses \citep[e.g.][]{Deming06,Huitson12,Wyttenbach15}.

We used HARPS archival data on HD 189733b, obtained under the programs 072.C-0488(E), 079.C-0828(A), and 079.C-0127(A). We chose not to use the data from 2006-07-29 because of the partial coverage of the transit (first half only) and because of the incomplete observation of the baseline. Instead, we used observations obtained during a total of three transits, which is the same dataset as was used by \cite{Wyttenbach15}, see their Table 1 for the observation logs. HARPS \citep{HARPS} is a fibre-fed, cross-dispersed high-resolution ($\mathrm{R} \sim 115000$) echelle spectrograph  at the ESO 3.6m telescope, which through its stability is ideal for measuring radial velocities. It covers wavelengths from $378$ nm to $691$ nm over 72 orders. The data were taken during the night of September 7, 2006, consisting of 28 exposures of 600 seconds each, and on the nights of July 19, 2007 and August 28, 2007, consisting of $38$ and $40$ exposures of $300$ seconds each, respectively. We retrieved the cross-correlation functions (CCFs) from the ESO\ archive. These CCFs are calculated by correlating the data with a binary mask, both computed for each order of the spectrograph and for the full wavelength range, which are standard products of the ESO archive. For comparison, we also used the values computed for the radial velocity using the full wavelength range and the errors on these values. We note that the data reduction as used by the ESO archive used different masks for different nights, that is, a K5 mask for the first night and a G2 mask for the others. We used the data for the first night reduced with the G2 mask with the HARPS Data Reduction Software by \cite{Wyttenbach15}.

\subsection{Radial velocity extraction from the orders}

To determine the radial velocity as function of wavelength for each observed spectrum, we used the CCFs determined for each order of the HARPS spectrograph. We first combined the CCFs by averaging them, inside wavelength passbands of $50$ nm chosen to be the same as used by \cite{Pont08} and \cite{Sing11}. These are shown in Table \ref{ldcoeff}.  In this way, we can keep the same limb-darkening coefficients as used in their studies and compare them directly. We note that the fourth and fifth passband overlap by half in wavelength because this is where the individual studies of \cite{Pont08} and \cite{Sing11} overlap. The second column of Table \ref{ldcoeff} shows the range of HARPS orders in each passband. There is no exact match with the passbands used by \cite{Pont08} and \cite{Sing11}, but differences are negligible. The orders were assigned to the different passbands based only on their central wavelength. HARPS order $97$ (central wavelength $\lambda = 631.06 \mathrm{nm)}$ was removed from our analysis because of strong telluric contamination from molecular oxygen absorption. 

In this way, we obtained seven CCFs per spectrum for the seven different passbands.  The radial velocity extraction was performed as in the HARPS Data Reduction Software, that is, by fitting Gaussians to the CCFs. The centre of the Gaussian is the RV of that passband. This approach of averaging the CCFs in a passband before determining the radial velocity is similar to the standard HARPS data reduction procedure, in which the CCFs from the different orders are first combined before they are fit with a Gaussian.

The photon-noise-limited uncertainty of a radial velocity measurement is discussed by \cite{Hatzes92}, who showed that
\begin{equation}
\sigma_{\rm{RV}} \propto S^{-0.5} \Delta \lambda^{-0.5} R^{-1.5},
\label{error_rv}
\end{equation} 
where $\sigma_{\rm{RV}}$ is the uncertainty in the radial velocity, $S$ is the flux, and $R$ is the spectral resolution. The variable $\Delta \lambda$ in practice is proportional to the number of spectral lines and their depths over the observed wavelength range. Since this is challenging to derive from first principles for the different passbands, we used the dispersion in the radial velocity time-series to scale the errors afterwards.

\section{Models for the Rossiter-McLaughlin effect}\label{sec:models}
We used the publicly available code AROME \citep{Boue13} to analytically compute the RM radial velocity anomaly as measured with the CCF method. This code, written in C, was given the orbital planetary and stellar parameters (the projected rotation velocity $Vsin(i)$, the semi-major axis $a$, the radius of the star $R_\mathrm{star}$, expressed in solar radii units, the inclination angle $i$, the mutual inclination angle $\lambda$, and the planet orbital period P, all taken from \cite{Triaud09}, see Table \ref{aromecoeff}), and the times of observation, which it subsequently used to calculate the CCFs and the expected RV measurements from a Gaussian fit to these. 
The fraction of flux blocked by a planet crossing the surface of its host star depends on the limb darkening of the star. For the calculation with AROME we chose the non-linear law of \cite{Claret00} to match those used in the analysis of \cite{Pont08} and \cite{Sing11}, that is,
\begin{equation}
\frac{I \left( \mu \right)}{I \left( 1 \right)} \ = \ 1 - \sum_{k=1}^{4} a_{k} \left( 1 - \mu^{k/2} \right)
\label{claret2000}
.\end{equation}

For the white-light RV time series we used the quadratic limb-darkening law coefficients as in \cite{Triaud09}, who studied the RM effect for this planet using the same data set. All the limb-darkening coefficients used in our analysis are given in Table \ref{ldcoeff}. Since we are interested in the dependence of the RM effect on the planet effective size, for each passband we selected the limb-darkening law corresponding to those wavelengths and calculated the anomaly for $5000$ values of the planet-star radius ratio ($R_{\mathrm{p}}/R_{\mathrm{star}}$), between $0.105$ and $0.205$ with a step of $2 \cdot 10^{-5}$. These boundaries were chosen arbitrarily with the purpose of including the measured planet-star radius ratio for this planet using the same HARPS data as \cite{Triaud09} ($0.1581 \pm 0.0005$). All other parameters were kept fixed.
 We note that for the analysis we are only interested in the relative change of the $R_{\mathrm{p}}/R_{\mathrm{star}}$ as function of wavelength, not in absolute values. Therefore it is not relevant to fit for parameters such as the projected rotation velocity V$\sin$i, the mutual inclination angle $\lambda$, the macroturbulence $z$, and the scaled semi-major axis $a/R_{\rm{star}}$ separately for each passband. These should be independent of wavelength and therefore can only result in a small vertical offset in the transmission spectrum, not in a change of slope. 

\begin{table}
\caption{Planet orbital and stellar parameters used to calculate the Rossiter-McLaughlin (RM) effect with the AROME code \citep{Boue13}. The orbital planetary and stellar parameters are taken from and \cite{Triaud09}. The remaining parameters, i.e. the line width of the non-rotating star $\beta_{0}$, the sigma of the Gaussian fit to the CCF $\sigma_{0}$, and the macro-turbulence parameter $z,$ were estimated by us and only have a marginal effect on the resulting RM anomaly.}% \protect\marginnote{text}\textcolor{red}{please
%remove these two sentences becaus table captions are only to be descriptive }\textcolor[rgb]{1,0.501961,0}{We note that since we are only interested in the change of the effective planet radius as function of wavelength, it is not relevant to fit any of these parameters separately for each passband. Since these are independent of wavelength, they can only result in a absolute change in $R_{\rm{p}}/R_{\rm{star\protect\marginnote{text}}}$}}

 % title of Table
%, the RV semi-amplitude of the star's reflex motion K, and the systemic RV $\gamma$
\label{aromecoeff}      % is used to refer this table in the text
\centering                          % used for centering table
\begin{tabular}{c c c c}        % centered columns (4 columns)
\hline\hline                 % inserts double horizontal lines
Parameter & value & units & reference \\

a & $0.03120$ & au & \cite{Triaud09}\\
$R_\mathrm{star}$ & $0.766$ & $R_\mathrm{\odot}$ &\cite{Triaud09} \\
i & $85.508$ & deg. & \cite{Triaud09}\\
$\lambda$ & $-0.85$ & deg. & \cite{Triaud09}\\
V$\sin (i)$ & $3.10$ & km/s & \cite{Triaud09}\\
%Vsin(i) & 3.316 & km/s & \cite{Triaud09}\\
$\beta_{0}$ & $1.3$ & km/s & $-$\\
$\sigma_{0}$ & $3.3$ & km/s & $-$\\
$z$ & $4.0$ & km/s & $-$\\
P & $2.218573$ & days & \cite{Triaud09}\\
%K & $0.202$ & km/s & \cite{Triaud09}\\
%$\gamma$ &$-2.2$ & km/s & \cite{Triaud09}\\
\hline
\hline                                   %inserts single line
\end{tabular}
\end{table}

\begin{table*}
\caption{Wavelength passbands, the corresponding HARPS orders, and limb-darkening coefficients as used in our analysis. The first five passbands and their non-linear limb-darkening coefficients were chosen to match those from \cite{Sing11}, while the last three match those from \cite{Pont08}. The coefficients for the entire HARPS range were taken from \cite{Triaud09} for a quadratic limb-darkening law. We note that order $97$ (central wavelength $\lambda = 631.06 \ \mathrm{nm}$) was strongly affected by telluric molecular oxygen absorption and was therefore removed from the analysis.}             % title of Table
\label{ldcoeff}      % is used to refer this table in the text
\centering                          % used for centering table
\begin{tabular}{c c c c c c}        % centered columns (4 columns)
\hline\hline                 % inserts double horizontal lines
Passband (nm) & Orders & c1 & c2 & c3 & c4 \\

%290-370 &  $0.5472$ & $-0.9237$ & $1.7727$ & $-0.4254$ &  \cite{Sing11} \\
370-420 & $146 - 161$  & $0.5836$ & $-0.8102$ & $1.7148$ & $-0.5389$  \\
420-470 & $131 - 145$ & $0.5089$ & $-0.4084$ & $1.3634$ & $-0.5302$  \\
470-520 & $118 - 130$ & $0.5282$ & $-0.3141$ & $1.1931$ & $-0.4994$ \\
520-570 & $108-117$ & $0.6158$ & $-0.3460$ & $1.0695$ & $-0.4578$  \\
550-600 & $104-111$ & $0.4621$ & $-0.2003$ & $0.9450$ & $-0.4045$ \\
600-650 & $95-96,98-102$ & $0.5148$ & $-0.2774$ & $0.9429$ & $-0.4033$  \\
650-700 & $90-93$ & $0.5620$ & $-0.3531$ & $0.9524$ & $-0.4096$  \\ 

\hline

%STIS & $90-161$ & $0.5598$ & $-0.4055$ & $1.2498$ & $-0.4945$\\
%ACS & $0.7601$ & $-0.5528$ & $0.9838$ & $-0.4226$ \\
%B & $0.6410$ & $-0.9192$ & $1.9071$ & $-0.7052$ \\
%V & $0.6564$ & $-0.7342$ & $1.6359$ & $-0.6963$ \\
Triaud & $90-161$ & $0.6355$ & $0.1488$ & \qquad & \qquad \\

\hline
\hline                                   %inserts single line
\end{tabular}
\end{table*}

\section{Data analysis}
The aim of our data analysis is to determine the change in the effective radius of the planet over the seven wavelength passbands. 
The first step in the analysis was to  remove the RV variations due to the reflex motion of the host star around the planet-star barycenter. In addition, common mode errors that are expected to be similar over the whole wavelength range covered by HARPS were removed, since they will not influence the change in $R_{\mathrm{p}}/R_{\mathrm{star}}$ as function of wavelength. The final step is the retrieval of $R_{\mathrm{p}}/R_{\mathrm{star}}$ for the seven passbands.

\subsection{Reflex motion of the star}
\begin{figure*}
\centering
\includegraphics[width = \hsize]{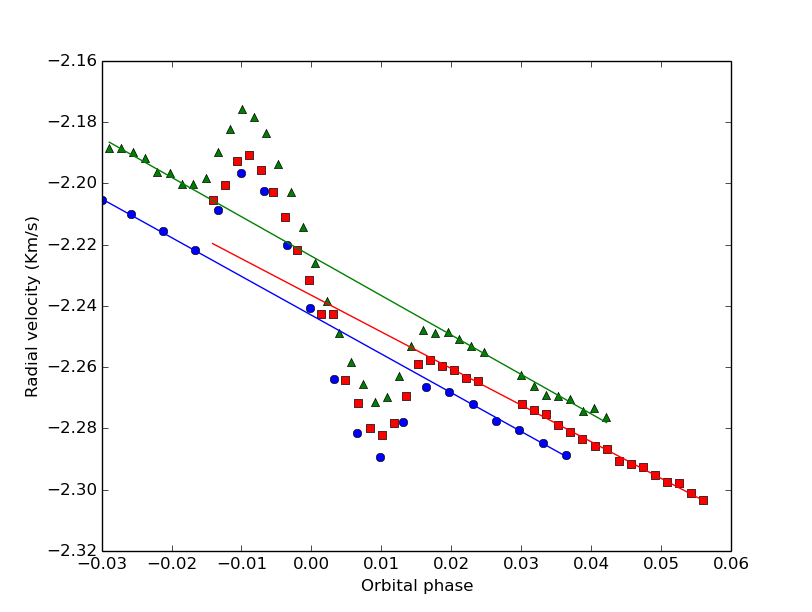} 
      \caption{White-light RV time series, plotted as blue dots for the first night, red squares for the second, and green triangles for the third night. The solid lines, with the same colours as used above, are the fitted reflex motion ($K$) and $v_{\rm{sys}}$ for each night. These values are shown in Table \ref{reflex_motion}.} \label{detrending}
\end{figure*}

Assuming that the orbit of HD 189733b is circular, the RV motion of the host star $v_{\rm{star}}$ can be modelled as a sinusoid with semi-amplitude $K$ and period $P$ given in Table \ref{aromecoeff}, plus a systemic radial velocity $v_{\rm{sys}}$. The exact values of $v_{\rm{sys}}$ differ from night to night because of the different stellar templates used to compute the CCFs (see Sect. 2). In addition, the RVs computed for the different passbands will also have small offsets because of the uncertainty in the reference RV for each order, see below. Stellar activity may also influence $v_{\rm{sys}}$ by spots covering parts of the rotating stellar disk. In addition, star spots rotating in and out of view may also result in small differences in $K$ between the nights \citep{Albrecht12}. We performed a least-squares fit of the out-of-transit RV data to obtain $K$ and $v_{\rm{sys}}$ for each night. These values are shown in Table \ref{reflex_motion} and in Fig. \ref{detrending}.

\begin{table}
\caption{Fitted values of the systemic velocity $v_{\rm{sys}}$ and semi-amplitude of the reflex RV motion of the star, $K$ for the three nights.} 
             % title of Table
\label{reflex_motion}      % is used to refer this table in the text
\centering                          % used for centering table
\begin{tabular}{r c c}        % centered columns (4 columns)
\hline\hline                 % inserts double horizontal lines
Observation & $v_{\rm{sys}}$ & $K$ \\
Night 1& $-2.2431 \pm 0.0002$ & $-0.202 \pm 0.003$ \\
Night 2&$-2.2366 \pm 0.0008$ & $-0.190 \pm 0.007$ \\
Night 3&$-2.2238 \pm 0.004$ & $-0.205 \pm 0.005$ \\
\hline
\hline                                   %inserts single line
\end{tabular}
\end{table}
The reflex motion was subsequently removed from the white-light RV time series and from that of each passband. Small residual velocity offsets at a level of a few hundred meters per second were still present in the different passband RV time-series (since different parts of the templates were used). These offsets were also fitted and removed. 

\subsection{Removal of common mode errors from the data}\label{sec:whiteres}
Common mode errors that are identical across all passbands because
of residual instrumental effects, star spots (to first order), stellar differential rotation, or other systematic effects \citep[e.g.][]{Triaud09}, can be removed from the data without harm. Therefore, we first performed a least-squares fit of the AROME models (Sect. 3) to the white-light RV time-series to determine the RM anomaly. This best fit was then removed from the white-light data to give the common mode errors on the time series, which were subsequently removed from the RV time series of each passband. To fit the white-light RV time series, a quadratic limb-darkening law was used with coefficients from \cite{Triaud09}, who analysed the same HARPS data as here. These coefficients are shown in Table \ref{ldcoeff}.

To determine the best value of the radius ratio $R_{\mathrm{p}}/R_{\mathrm{star}}$ for the white-light RV time series, we computed the chi-squared $\chi^{2}$ for each model to the data, with the uncertainties rescaled such that the reduced $\chi^{2}$ of the best-fit model is equal to unity. The $1\sigma$-confidence interval was then determined by selecting the values of the radius ratio at which $\Delta \chi^{2} =1$.

\begin{figure}
   \centering
   \includegraphics[width=\hsize]{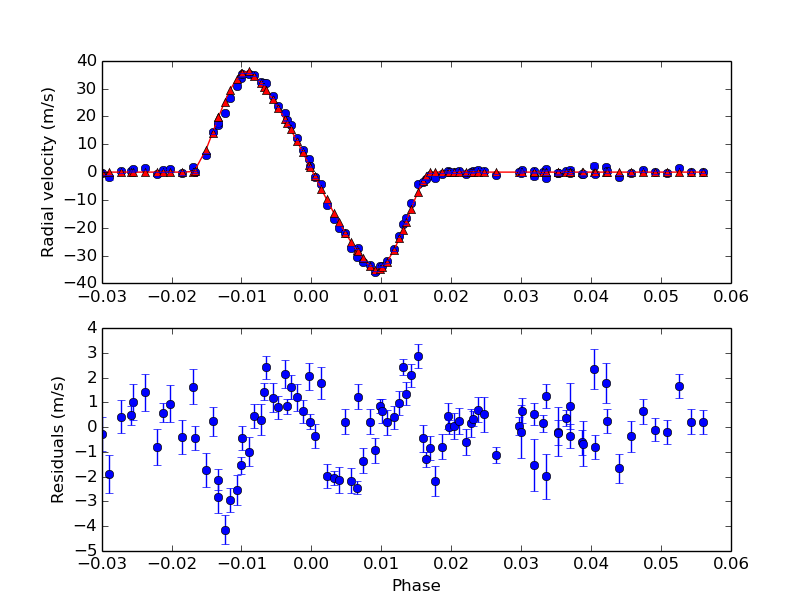}
      \caption{Top panel: the white-light RV time-series with the best fit to the Rossiter-McLaughlin anomaly. The fit yields a planet/star radius ratio of $R_{\mathrm{p}}/R_{\mathrm{star}} = 0.1585 \pm 0.0007$, in agreement with the measurement of \cite{Triaud09}. In the bottom panel the residuals to the fit are shown. These residuals contain information about possible star spots crossed by the planet during a transit and possible effects due to stellar differential rotation or macroturbulence. These residuals were subsequently removed from the radial velocity time-series of the individual passbands.}
\label{whitefit}
   \end{figure}

The best fit, shown in Fig. \ref{whitefit}, yielded a planet/star radius ratio of $R_{\mathrm{p}}/R_{\mathrm{star}} = 0.1585 \pm 0.0007$, in agreement with the value found by \cite{Triaud09} for this planet based on the same HARPS data ($0.1581 \pm 0.0005$).

\subsection{Fit of the Rossiter-McLaughlin curve}
\begin{figure*}
\begin{tabular}{cc}
\includegraphics[width = 0.5\hsize]{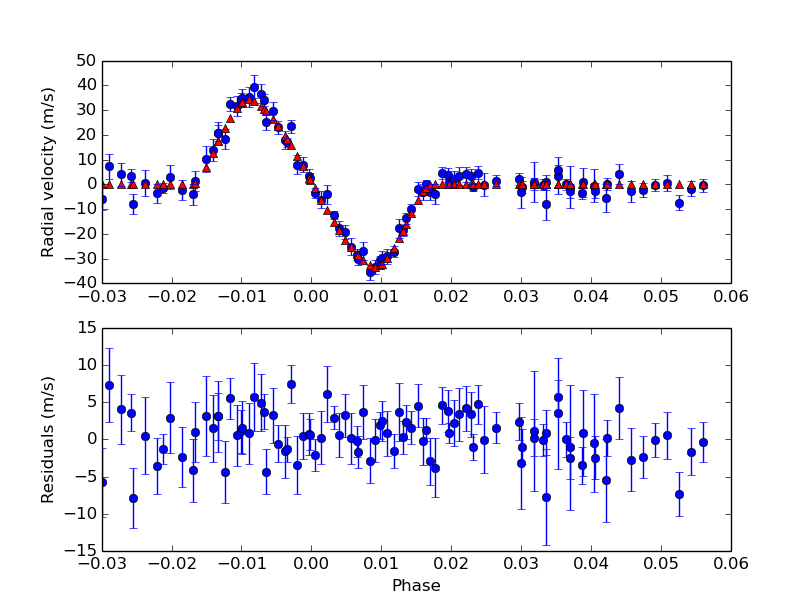} & \includegraphics[width = 0.5\hsize]{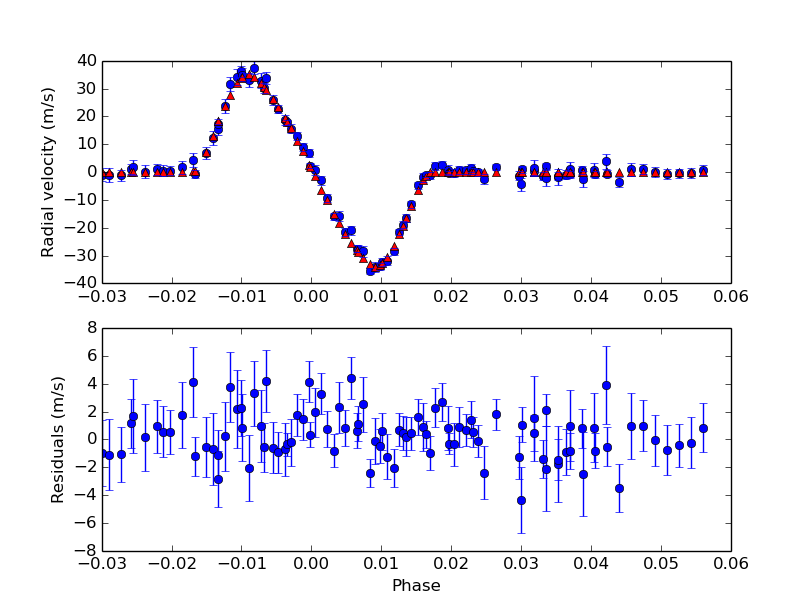} \\

\includegraphics[width = 0.5\hsize]{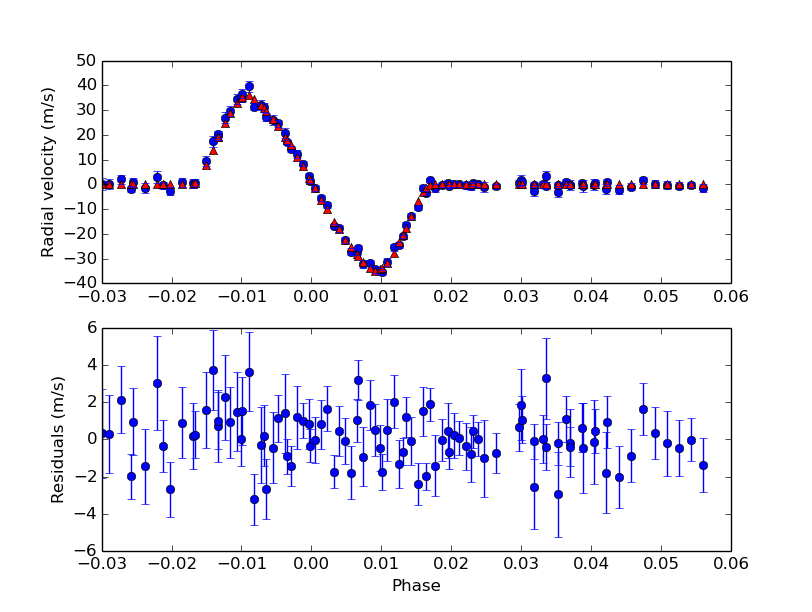} & \includegraphics[width = 0.5\hsize]{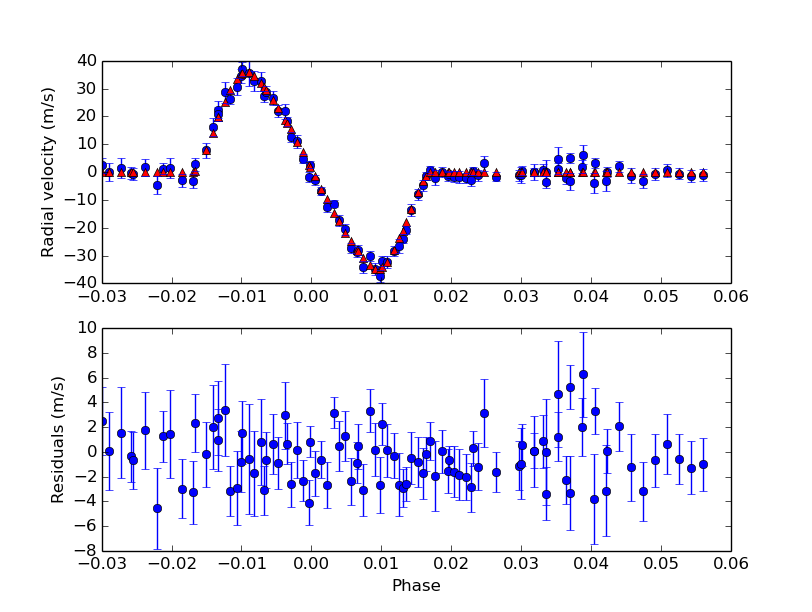}\\

\end{tabular}
      \caption{Fit of the Rossiter-McLaughlin effect for all the individual wavelength bins (from top left to bottom right). In the top panels of each figure, the RVs are shown in blue, computed for each bin and in red the fit to them; in the bottom panels of each figure the residuals of the fit are shown.}
         \label{fit-rm}
\end{figure*}

\addtocounter{figure}{-1}
\begin{figure*}
\begin{tabular}{cc}
\includegraphics[width = 0.5\hsize]{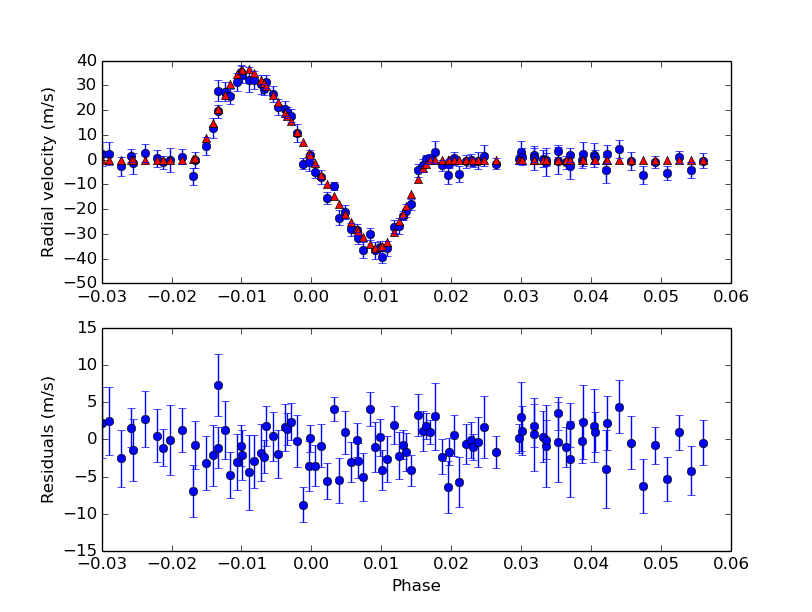} & \includegraphics[width = 0.5\hsize]{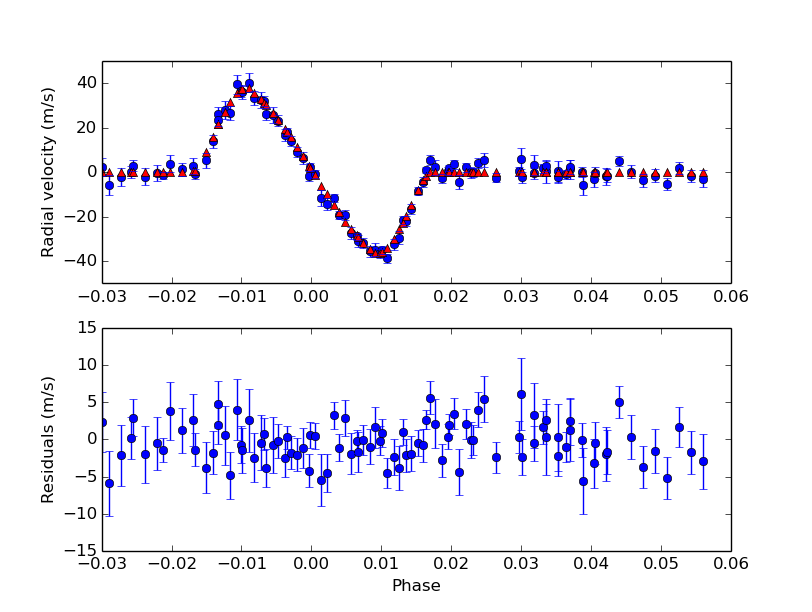}\\
\includegraphics[width = 0.5\hsize]{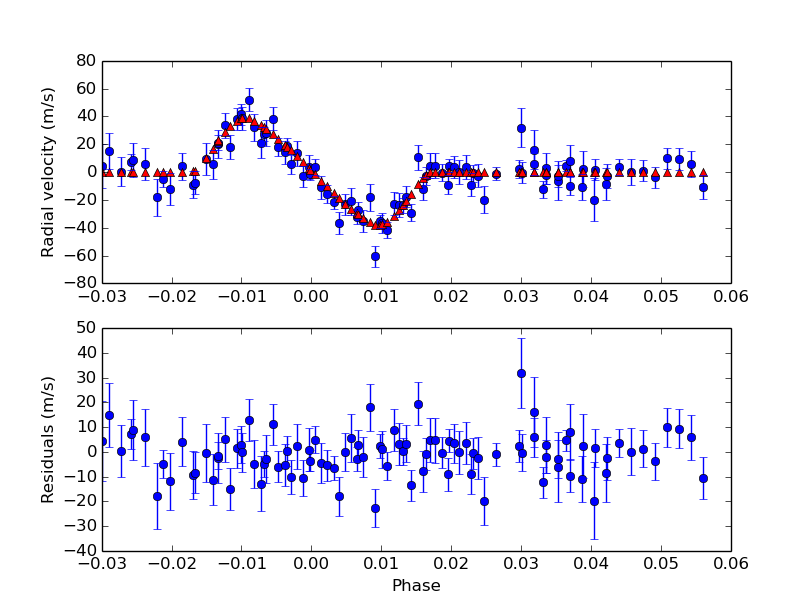} &\quad \\
\end{tabular}
\caption{Continued...}
\end{figure*}

The resulting time series were fitted using the grid of models described in Sect. \ref{sec:models} by $\chi^{2}$ minimization. The uncertainties on the measurements were rescaled such that for the best fit $\chi_{\rm{reduced}}^{2} = 1$, and the associated $1\sigma$ errors were determined from the condition that $\Delta \chi^{2} = 1$. We note that even if the limb-darkening coefficients chosen for each passband were wrong by a quarter of a passband width, this would only contribute to the uncertainty on the retrieved $R_{\mathrm{p}}/R_{\mathrm{star}}$ value on a level of less than  $0.5 \sigma$.  

\begin{table}
\caption{$R_{\mathrm{p}}/R_{\mathrm{star}}$ values obtained for the different wavelength passbands. The quoted uncertainties are for the relative values, not for absolute measurements. The larger error at the longest wavelength arises because there are only four spectral orders contributing to that passband. The errors are for a $1\sigma$ confidence level.}             % title of Table
\label{rprstar}      % is used to refer this table in the text
\centering                          % used for centering table
\begin{tabular}{c l }        % centered columns (4 columns)
\hline\hline                 % inserts double horizontal lines
Passband (nm) & $R_{\mathrm{p}}/R_{\mathrm{star}}$ \\

%290-370 &  $0.5472$ & $-0.9237$ & $1.7727$ & $-0.4254$ &  \cite{Sing11} \\
370-420 &  $0.1626 \pm 0.0015$ \\
420-470 &  $0.1610 \pm 0.0009$ \\
470-520 &  $0.1606 \pm 0.0007$ \\
520-570 &  $0.1582 \pm 0.0011$ \\
550-600 &  $0.1591 \pm 0.0015$ \\
600-650 &  $0.1595 \pm 0.0012$ \\
650-700 &  $0.1618 \pm 0.0035$ \\ 
\hline
\hline                                   %inserts single line
\end{tabular}
\end{table}

\section{Results} \label{result}
\begin{figure}
   \centering
   \includegraphics[width=\hsize]{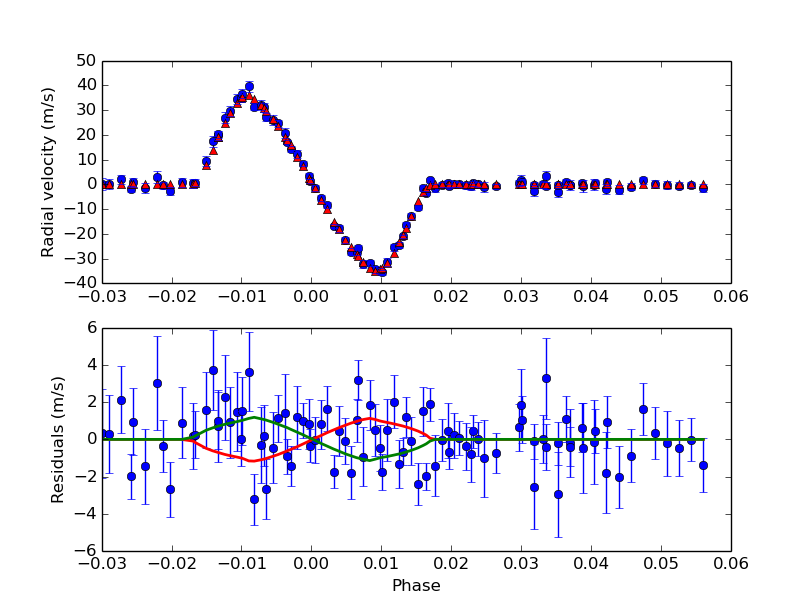}
      \caption{Upper panel: the radial velocity time-series for the fifth passband (550-600 nm) and the corresponding best-fit model (blue dots and red line, respectively). The lower panel shows the residuals to the fit, together with the 5$\sigma$ upper and lower limit of the $R_{\mathrm{p}}/R_{\mathrm{star}}$ ratio (solid red and green lines, respectively). }
\label{sensitivity}
   \end{figure}
The least-squares fits to the RM effect in the individual passbands are shown in Fig. \ref{fit-rm}. The scatter of the residuals ranges from $2 \ \rm{m}/\rm{s}$ for the 470-520 nm passband to $9 \ \rm{m}/\rm{s}$ for the 650-700 passband. The $R_{\mathrm{p}}/R_{\mathrm{star}}$ obtained for the different wavelength passbands are given in Table \ref{rprstar} and are shown in Fig. \ref{rayleigh-fit}. In Fig. \ref{sensitivity} we show the data of the fifth passband with the best-fit model, and that for 5$\sigma$ smaller and larger planet/star radius ratios to illustrate the sensitivity of our fitting method.

Star spots, unocculted by the planet, change the effective size of the star as function of wavelength. Therefore, they can have an equivalent effect on $R_{\mathrm{p}}/R_{\mathrm{star}}$ as a change in the effective size of the planet, which is what we wish to measure with transmission spectroscopy. \cite{Sing11} corrected for unocculted star spots by assuming a spot temperature of  $T_{\rm{spot}} = 4250 \pm 250 \ \rm{K,}$ causing a stellar flux reduction of $1 \%$ at $600$ nm. We performed that same analysis. However, because our uncertainties in $R_{\mathrm{p}}/R_{\mathrm{star}}$ are an order of magnitude larger, this spot correction contributes only at the 0.1 $\sigma$ level and was therefore ignored.

\subsection{Rayleigh-scattering slope}

As in \cite{Sing11}, \cite{Pont08}, and \cite{Pont13}, we interpreted the optical transmission spectrum of HD189733b as due to Rayleigh scattering. The Rayleigh scatter cross-section can be written as $\sigma = \sigma_{0} \left(\lambda/\lambda_{0} \right)^{\alpha}$, with $\alpha = -4$, and therefore the slope of the planet radius as a function of wavelength is given by \cite{Cavelier}
\begin{equation}
\frac{\mathrm{d} R_{\mathrm{p}}}{\mathrm{d} \mathrm{ln} \lambda} = \frac{\mu g}{k} \alpha T
\label{rayleigh}
,\end{equation}
where $\mu$ is the mean mass of the atmospheric particles, taken as $2.3$ times the mass of the proton, $k$ is the Boltzmann constant, and $g$ the surface gravity. From the slope of the fit to the $R_{\mathrm{p}}/R_{\mathrm{star}}$ ratio as a function of the natural logarithm of the wavelength, an estimate of the atmospheric temperature (at pressures where the scattering takes place) can be derived. 

\begin{figure}
   \centering
   \includegraphics[width=\hsize]{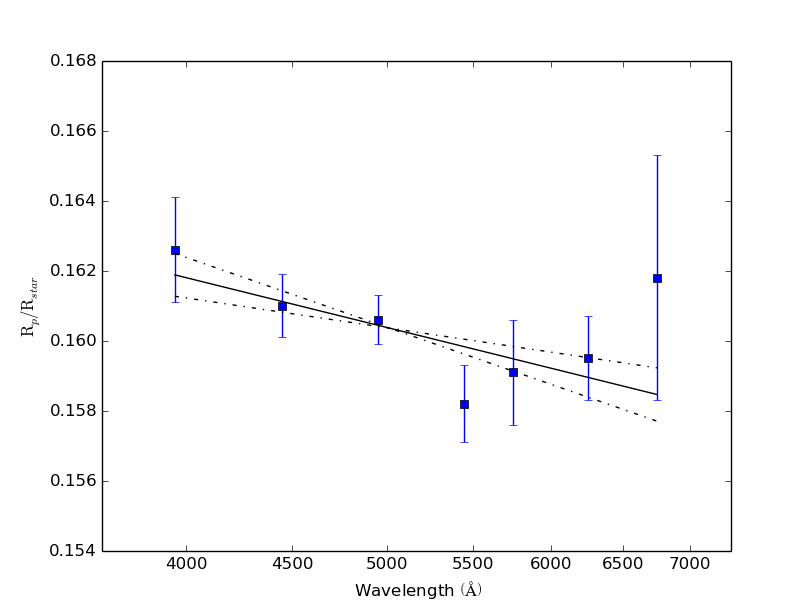}
      \caption{Resulting planet-star radius ratio $R_{\mathrm{p}}/R_{\mathrm{star}}$ as function of wavelength, indicating the 1$\sigma$ error bars. The solid line indicates the least-squares fit of a Rayleigh-scattering slope to the data, using Eq. \ref{rayleigh}. The dashed lines indicate the 1$\sigma$ error margin on the slope. Note that the abscissae axis is in a logarithmic scale. The linear fit yields a gradient of $-0.0064 \pm 0.0026$ ln[$\AA$]$^{-1}$, corresponding to a temperature of $T = 2300 \pm 900 \mathrm{K}$.}
\label{rayleigh-fit}
   \end{figure}

We performed a least-squares fit for the data as shown in Fig. \ref{rayleigh-fit}, weighting down the data points of passbands $520-570$ nm and $550-600$ nm by a factor 0.75 because they overlap by half. The best fit is shown in Fig. \ref{rayleigh-fit} and coresponds to a gradient of $-0.0064 \pm 0.0026$ ln[$\AA$]$^{-1}$, which is significant at a $2.5 \sigma$ level. Assuming $R_\mathrm{star} = 0.766 R_\mathrm{\odot}$ \citep{Triaud09}, this gradient corresponds to an atmospheric temperature of $T = 2300 \pm 900 \mathrm{K}$ (errors at $1\sigma$ confidence level).

\begin{figure}
   \centering
   \includegraphics[width=\hsize]{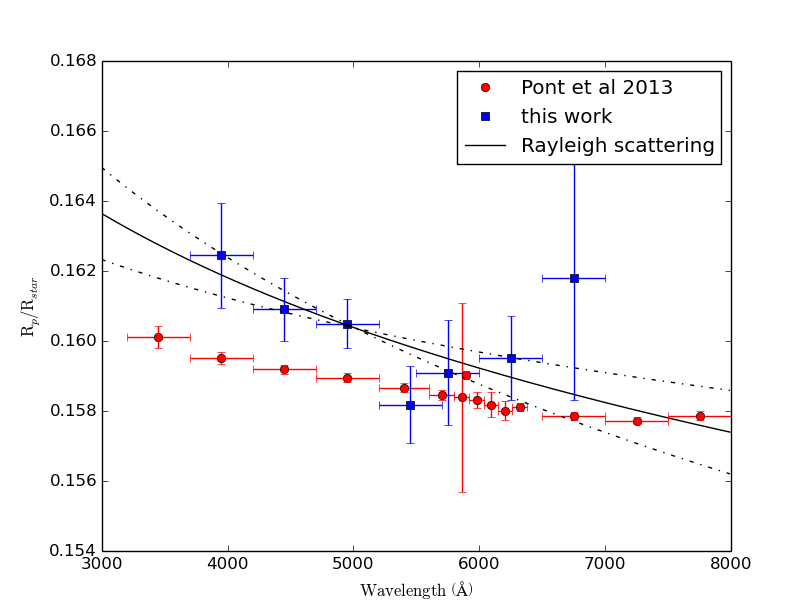}
      \caption{Comparison of our $R_{\mathrm{p}}/R_{\mathrm{star}}$ measurements (blue squares) with those of \cite{Pont13} (red circles). The 1$\sigma$ error bars are indicated. The solid and dashed lines are the same as in Fig. \ref{rayleigh-fit}.}
\label{final-fit}
   \end{figure}

We compared our results with those of \cite{Pont13}, who performed the combined analysis of the data in \cite{Sing11} and \cite{Pont08}. This is shown in Fig. \ref{final-fit}. We note that we shifted the data from \cite{Pont13} up by 0.0002 in $R_{\mathrm{p}}/R_{\mathrm{star}}$ to match our values. Our analysis is not meant to give precise absolute values of $R_{\mathrm{p}}/R_{\mathrm{star}}$ (which is influenced by the choices in stellar parameters such as projected rotational velocity and macro turbulence), only the relative change as function of wavelength. The uncertainties in our measurements are typically an order of magnitude larger than those of \cite{Pont13}. 

\cite{Pont13} inferred two different atmospheric temperatures for two separate wavelength regimes, below and above $550 \mathrm{nm}$. Above $550 \mathrm{nm}$ they found a gradient that corresponds to a temperature of about $1300 \mathrm{K}$, while at shorter wavelengths the slope steepened corresponding to a temperature of about 2000 K. As can be seen in Fig. \ref{final-fit}, the uncertainty in our data is too large to fit multiple slopes. The uncertainty in temperature found in our analysis agrees with both temperatures as found by \cite{Pont13}. 
The temperature in the upper atmosphere of HD 189733b has also been estimated usign the NaI doublet line cores. \cite{Huitson12} used HST data and found a temperature of $2800 \pm 400 \ \mathrm{K}$, while \cite{Wyttenbach15} found a temperature of  $2500 \pm 400 \ \mathrm{K}$, using the same HARPS data set.


\subsection{Other interpretations of the slope}
In our analysis we assumed that the slope in the transmission spectrum of HD189733 b is due to Rayleigh scattering, as was also assumed by \cite{Pont08}, \cite{Cavelier}, \cite{Sing11}, and \cite{Pont13}.
% \marginnote{text}\textcolor{red}{please integrate
%the list of references properly into the text by changing the
%semicolons to commas and adding "and" before the last reference} \cite{Pont08,Cavelier,Sing11,Pont13}. 
However, in recent literature it has been advocated that star spots or plages might also result in the observed slope in the transmission spectrum. \cite{McCullough14} reinterpreted the available measurements with a clear planetary atmosphere. They found that an unocculted spot fraction of 4\% can mimic the increase in effective planet size towards shorter wavelengths. Occultation of stellar plages has also been claimed to be able to give rise to the slope in the spectrum of HD 189733b by \cite{Oshagh14}. We note that the chromatic RM effect as used in our analysis to determine the effective planet radius as function of wavelength is affected by star spots and/or plages in a similar fashion as classical transmission spectroscopy. Therefore, this analysis cannot contribute to this discussion at this stage. 

However, it is worth noting that an increase in planetary radius towards UV wavelengths has been detected for a number of planets, for instance, GJ3470b \citep{Nascimbeni13,Biddle14}, WASP-6b \citep{Nikolov15}, and WASP-31b \citep{Sing15}. Interestingly, the host stars of all these planets and that of HD189733b are active stars. This might indicate that unocculted star spots or plages may be the cause of this effect. \cite{Oshagh14} also
advocated this for the particular case of GJ3470b. A more detailed investigation is required to understand the relationship between stellar activity and slopes in transmission spectra of orbiting planets.

\subsection{Future observations with ESPRESSO}

We showed that with three transit observations of HD189733b with HARPS, we obtained a relative precision of the planet radius as a function of wavelength, which is about an order of magnitude lower than that achieved for five transits with STIS at the HST \citep{Pont13}. Since the chromatic RM technique is photon limited, a significant increase in precision can be expected from the forthcoming ESPRESSO spectrograph at ESO's Very Large Telescope \citep{ESPRESSO}. 

\begin{figure}
   \centering
   \includegraphics[width=\hsize]{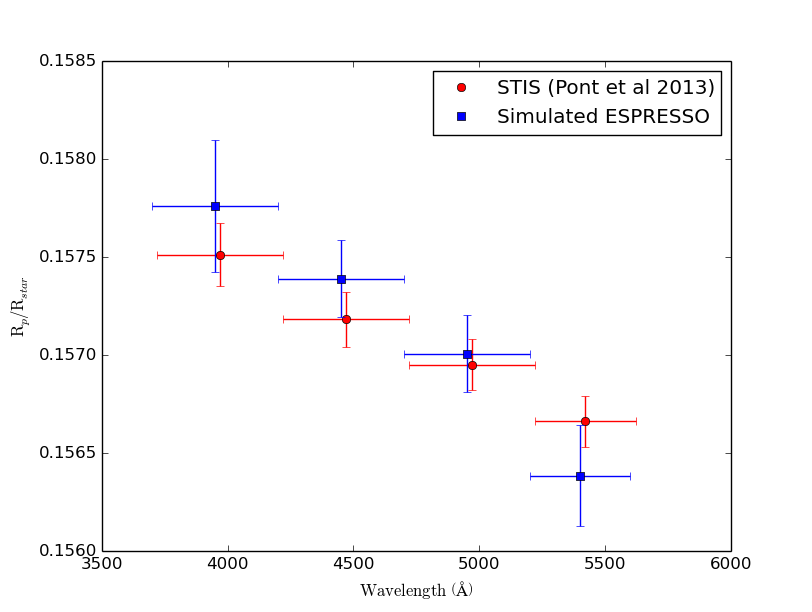}
      \caption{Simulated $R_{\mathrm{p}}/R_{\mathrm{star}}$ for five transits observed with the future ESPRESSO spectrograph on the VLT (blue squares). Observations of five transits by the STIS spectrograph on HST as presented by \cite{Pont13} are indicated by red circles. These are shifted slightly redward for clarity. All error bars indicate 1$\sigma$ uncertainty intervals. These are larger by between 1.5 and 2 times for the ESPRESSO than for the STIS observations.                                 }
\label{espresso-fit}
   \end{figure}

We simulated future ESPRESSO observations by scaling the current accuracy achieved with HARPS by the increase in collecting area of the telescope and the increase in throughput of the ESPRESSO spectrograph, also taking into account the slight increase in spectral resolution to $R=120000$.  The photon-noise contribution to the radial velocity error is, according to Eq. \ref{error_rv}, $\sigma_{RV} \propto S^{-0.5} R^{-1.5}$, where $S$ is the received flux from the star, and $R$ the resolving power. Figure \ref{espresso-fit} shows the results of our simulation for five transits observed with ESPRESSO (blue squares) with their associated 1$\sigma$ uncertainties. In red circles are shown the data points as presented by \cite{Pont13} for also five transits observed with STIS. The expected uncertainties for ESPRESSO are about 1.5-2 times larger than for STIS. This is particularly important, taking into account the limited and uncertain future of the Hubble Space Telescope, and the fact that the future James Webb Space Telescope will not cover this wavelength regime.

We do note that the amplitude of the RM effect depends on the spin of the star and its angle with the planet's orbit. Therefore, not all transiting planetary systems, for instance those orbiting slow rotators, are equally favourable for this technique.

\section{Conclusion}
We used the chromatic Rossiter-McLaughlin technique to probe the slope in optical transmission spectrum of HD189733b using archival HARPS data.
 We showed that this technique can be used to measure this slope and found a change in radius ratio $R_{\mathrm{p}}/R_{\mathrm{star}}$ of $-0.0064 \pm 0.0026$ ln[$\AA$]$^{-1}$ (2.5$\sigma$). Assuming this is due to Rayleigh scattering, it corresponds to an atmospheric temperature of  $T = 2300 \pm 900 \mathrm{K}$.

The precision achieved with HARPS per transit is about an order of magnitude lower than that with STIS on the Hubble Space Telescope. This method will be 
particularly interesting in conjunction with the new echelle 
spectrograph ESPRESSO, which currently is under construction for ESO’s Very Large Telescope and will provide a gain in signal-to-noise
ratio of about a factor $\text{}$4 compared to HARPS. This will be of great value because of the limited and uncertain future of the Hubble Space Telescope and because the future James Webb Space Telescope will not cover this wavelength regime.

\begin{acknowledgements}
 This work is part of the research programmes PEPSci and VICI 639.043.107, which are financed by the Netherlands Organisation for Scientific Research (NWO). Based on observations made with ESO telescopes at the La Silla Paranal Observatory under programme 072.C-0488(E), 079.C-0828(A), and 079.C-0127(A). We thank C. Lovis, A. Wyttenbach, D. Ehrenreich, and F. Pepe for useful discussion and insights. 
\end{acknowledgements}
\bibliographystyle{aa}
\bibliography{references-harps}

\end{document}